\documentclass[usenatbib]{mn2e}
\usepackage{times}
\usepackage{epsf}
\usepackage{natbib}
\usepackage{fixltx2e}
\usepackage{cite}
\usepackage{rotating}
\newcommand{\etal}{{et al.}}

\begin {document}
\title[The dynamics and environmental impact of 3C452]{The dynamics and environmental impact of 3C\,452}
\author[D.L.\ Shelton \etal
]{D.L.\ Shelton$^1$\thanks{E-mail:D.Shelton@herts.ac.uk},
  M.J.\ Hardcastle$^1$ and J.H.\ Croston$^2$ \\
$^1$ School of Physics, Astronomy \& Mathematics, University of Hertfordshire, College Lane, Hatfield, AL10 9AB\\
$^2$ School of Physics and  Astronomy, University of Southampton, Southampton, Hampshire, SO17 1BJ}
\maketitle
\begin{abstract}
We present a detailed analysis of a new {\it XMM-Newton} observation
of the FRII radio galaxy 3C\,452 and its environment. We detect X-ray
emission from the hot intragroup medium and measure its temperature as
well as obtaining the surface brightness and pressure profiles. We
provide evidence that 3C\,452 is currently heating its environment, measuring a temperature of $1.18\pm0.11$ keV for the immediate environment of the radio source compared to $0.86^{+0.13}_{-0.05}$ keV for the outer environment) . We
also present evidence that the outer regions of the lobes are
overpressured (internal pressure of $2.6\times10^{-13}$ Pa and external pressure of $1.11\pm{0.11}\times10^{-13}$ Pa at the edge of the lobes) and therefore are driving a shock at the lobe edges (with a temperature which we constrain to be $1.7^{+0.9}_{-0.5}$ keV),
while the inner regions of the lobes are underpressured and
contracting. Taking into account the very large amount of energy
stored in the lobes, we show that this relatively low-powered FRII radio galaxy will have an
extremely significant impact on its group environment.
\end{abstract}


\section{Introduction}

Understanding how a powerful radio galaxy interacts with a poor
environment is important in exploring the impact of radio galaxies on
the properties of the group population. The most powerful class of jet-driven AGN
outbursts that we know of are the FRII radio galaxies \citep{fr74}. X-ray observations of some of these objects suggest that
the energy input from the expanding radio galaxy can be similar to the
binding energy of the local intergalactic medium (e.g. \citealt{kbhecwm07}, \citealt{hck07}); thus it is clear that in principle those
sources will have a profound effect on their environments over their
lifetimes.

The radio lobes in FRII radio galaxies are the result of the
interactions between the relativistic electrons that have passed up
the jet and the external medium, where the electrons gain energy at
the jet termination shocks and expand out into the external medium
\citep{s74}. It has been known for many years that the inflation of
these lobes should affect the surrounding intergalactic medium (IGM).
Even when the lobes expand at sub-sonic speeds (where the internal
lobe and ram pressures are comparable to the external pressure), they
do $p{\rm d}V$ work on the hot gas. When the lobes expand at supersonic speeds, we expect additional work to be done to shock-heat the gas. This is a non-adiabatic process that effects the entropy as well as the internal energy of the external medium.

However, it is only recently that we have been able to make
detailed statements about the internal and external pressures of FRII
radio galaxies in typical environments, which determine their
dynamics. This change has come about because the current generation of
X-ray telescopes, {\it XMM-Newton} and {\it Chandra}, have the ability
to detect inverse-Compton emission from the lobes together with
thermal emission from their hot gas environments, which are often
relatively poor (\citealt{cbhw04}; \citealt{kbhecwm07}; \citealt{hec07}; \citealt{bwhc07}). Detection and characterization of
these X-ray components allows us to describe the current dynamics
of the radio lobes and gives us some idea what the energy input to
the external environment over the lifetime of the radio source must
have been. The impact that the AGN outbursts have on the hot IGM has
implications for the history and evolution of the hot phase of the
baryonic matter and the nature of `feedback' from radio-loud active
galaxies. When discussing the evolution of hot gas, it is important to
understand that, although powerful radio-loud AGN are rare, their
impact can be long lived: a single FRII outburst can
have an effect on the IGM surrounding the AGN galaxy that exceeds that of a
large number of less violent, longer-lived low-power radio sources
\citep{ba03} and many groups that no longer show signs
of AGN may once have hosted FRII sources.

Detailed studies of the dynamics of radio galaxies using X-ray
observations rely on the capability of separating the thermal and
non-thermal emission from the source. Some of the non-thermal emission
comes in the form of synchroton and inverse-Compton emission from the
jets and hotspots, which is generally best studied at high resolution
with {\it Chandra}. However, the dominant non-thermal component in
powerful radio galaxies is inverse-Compton emission from the lobes.
This occurs when electrons in the lobes scatter cosmic microwave
background (CMB) photons to X-ray energies (e.g. \citealt{flmkf95};
\citealt{tkmiiikty98} \citealt{hbchlw02}; \citealt{ks05};
\citealt{chhbbw05}). To be able to separate the non-thermal and
thermal components, we need to study objects where we can obtain a
significant number of counts from both the lobes and the thermal
environment. In particular, to avoid the inverse-Compton emission
being dominated by thermal bremsstrahlung (cf. \citet{hc10}) we need to select objects that are powerful, close and inhabit
relatively poor environments.

In this paper we study the nearby, powerful radio galaxy 3C452. 3C452
is an FRII radio galaxy which has a symmetrical double-lobe morphology
and a relatively  uniform surface brightness distribution \citep{bblprs92}; its
largest angular size is $\sim 5$ arcmin, which at its redshift of
0.0811 corresponds to a projected linear size of $\sim 450$ kpc.
Images from the {\it Hubble Space Telescope} and various sky surveys
(such as 2MASS) suggest that the environment is a poor group, with a clear excess of galaxies within the instrument's field of
view; however, group membership has not been confirmed
spectroscopically for any of the nearby galaxies.

\citet{itmismma02} discussed {\it Chandra} observations of 3C\,452,
showing that there is clear inverse-Compton emission from the lobes.
However, their observations were not sensitive enough to make
measurements of the properties of any extended thermal environment. In
this paper we describe the results of a deep {\it XMM-Newton}
observation of 3C452. {\it XMM-Newton's} higher sensitivity and larger
field of view when compared to {\it Chandra} allow us to detect and
constrain the properties of the group-scale gas around 3C452 and to
search for evidence both for shock heating and for other work done on
the external medium by the radio galaxy.

In the cosmology used in this thesis, the luminosity distance to 3C452 is 369 Mpc, and 1 arcsec corresponds to 1.53 kpc. Errors quoted on the results of X-ray
fitting are 90 per cent confidence errors.

\section {Observation \& Data Reduction}
\label{observations}

We observed 3C452 with the EPIC instrument on board {\it XMM-Newton}.
The exposure time for the MOS1 and MOS2 cameras were 69.587 ks and
69.621 ks respectively. For the PN, the exposure time was 54.024 ks.
The observation began on 2008 November 30.
 
The datasets were processed using the Scientific Analysis Software (SAS) version 9.0
and the standard pipeline tasks {\it emchain} and {\it epchain}.
The data was filtered using PATTERN $<=4$ (MOS) or PATTERN $<=12$ (PN)
and we used the bit-mask flags 0x766a0600 for the MOS camera and
0xfa000c for the PN camera. These are equivalent to the standard
flagset \#XMMEA\_EM/P, but they also include out of field-of-view
events and exclude any bad columns and rows found in the data. We did
not need to filter the data in the time domain, as there were not any
background flares in the dataset. The dataset was also energy filtered
between 0.2 and 12 keV for the MOS cameras and between 0.2 and
15 keV for the PN camera, using {\it xmmselect}. We then used {\it
  evigweight} to apply a vignetting correction to the files and
extracted spectra from each camera using {\it evselect}. {\it
  Evigweight} is a standard SAS tool which properly propagates
statistical errors. We use the SAS tool {\it arfgen} to create the
ancillary response files. Vignetting is not taken into account when creating the ancillary reponse files as it is already accounted for by the weighting factor applied by {\it
evigweight}.

In order to identify suitable regions for spectral analysis, it is first necessary to create an
image. We used {\it xmmselect} to create an image for each camera
in the 0.3--8 keV energy band, and then combined these images to generate an exposure map for each camera to correct for chip gaps.

We used a {\it Chandra} observation, first described  by \citet{itmismma02}, to identify point sources in the image. These were then removed from the combined image and Gaussian smoothing was used. We experimented with a range of kernel sizes so structure could be emphasized on different size scales (see Fig.\ \ref{image}). Figure \ref{lobes} shows the X-ray emission from the lobes of 3C\,452.

Finally, we made use of a radio map of 3C\,452 from the online 3CRR
Atlas\footnote{http://www.jb.man.ac.uk/atlas/}, at a frequency of 1.4
GHz and with a resolution of 6 arcsec.

\section{Spectroscopic Analysis}

In this section we discuss the two methods we used to extract spectra
for the various components of the radio galaxy and its environment and
then go on to discuss our results. Throughout the spectroscopic
analysis, we used events in the energy range 0.3 -- 8 keV.

\subsection{Background methods}
\label{background}

We used two approaches to determining the appropriate background for
our spectra: local background regions and a full double-subtraction
technique that takes account of the spatially variable particle
background. Where possible we used a local background for simplicity.
However, doing this for regions over which the particle background
might be spatially variable leads to bias in the results, and so for
larger regions we used closed-filter data with instrument modes matching our observations to constrain the particle
background properties. In this situation, we used the method of Maurin \etal\ (in prep); which is described by \citet{chbwl08}. As
in \citet{chbwl08}, we calculated particle background scaling
factors by comparing the 10--12 keV (MOS) and 12--14 keV (PN) count
rates for the source and closed-filter datasets\footnote{http://xmm2.esac.esa.int/external/xmm\_sw\_cal/background/filter\_closed/index.shtml}. The exposure times for the data sets were 107 ks for MOS1, 83 ks for MOS2 and 196 ks for PN. We used a circular region with a radius of 12,000 pixels in the energy ranges stated above. After this we extracted spectra for the source files and the background files and measured the count rate. Our scaling factors (ratio between the count rate of the source files and the count rate for the closed filter files) for the MOS1, MOS2 and PN cameras are 1.54, 1.70 and 1.63 respectively.
\\
\indent We chose a background region which was near the source, but did not contain any emission from it (this region lies between 320 and 480 arcseconds from the source as shown in Section \ref{spectralregion}) The spectra for the X-ray background for the three cameras,
determined from this background region, were then modelled in
{\sc xspec}, ignoring the energy range between 1.4 and 1.6 keV
where an instrumental line affects the fit. The X-ray background is
modelled as a double {\it apec} + absorbed power law, with the column
density fixed to the Galactic value and abundance fixed at solar. The two {\it apec} models account
for the emission from the Galactic bubble and the power law accounts
for the cosmic X-ray background. The temperatures were allowed to
vary, as were the normalizations, but the power law index was fixed at
$\Gamma = 1.41$ \citep{lwpdl02}. The normalizations were allowed to vary to help minimize intrumental differences between the MOS and PN cameras and to account for differences in the local X-ray background in different directions.  Finally, a scaled version of this
spectrum, with all parameters fixed, was used in fits to the spectra
for each source region. Table \ref{background-fit} shows the parameters of
the best-fitting model for the background region we used. Our model is in agreement with the model of \citet{lwpdl02}.

\begin{figure*}
\epsfxsize 12.5cm
\epsfbox{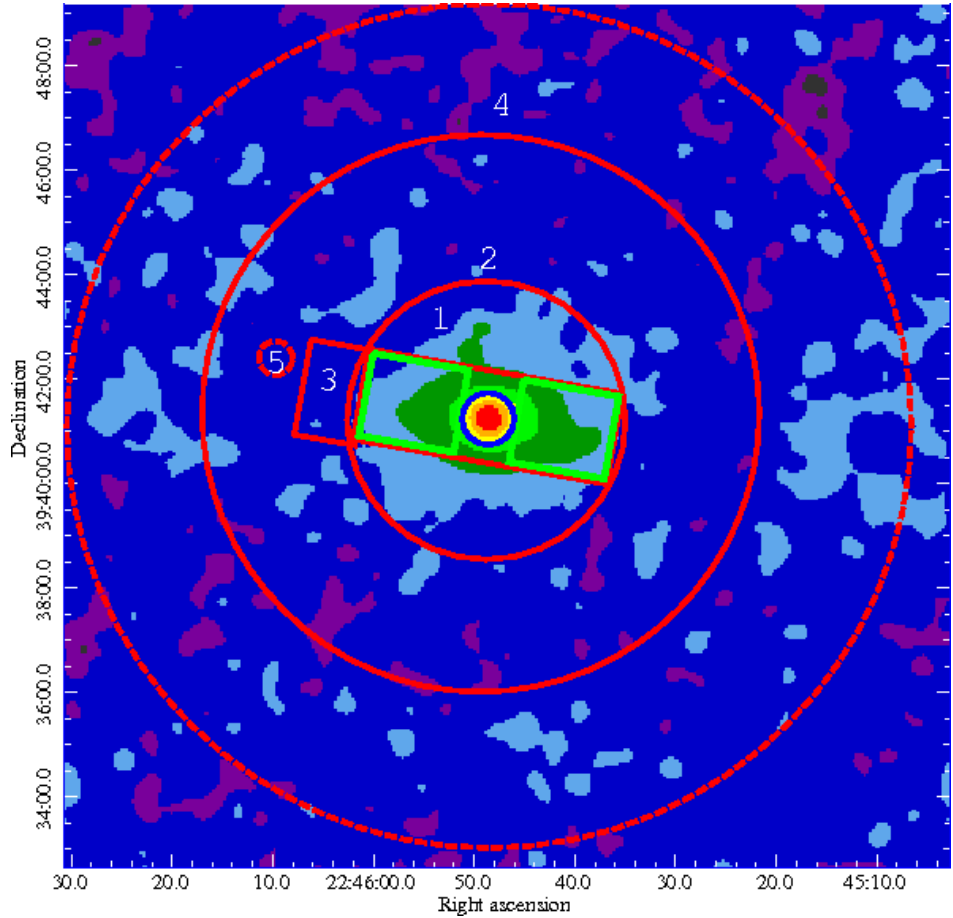}
\caption{A smoothed image of 3C452 in the energy band 0.3 - 8 keV
  after the removal of point sources. The image has been binned to 3.2
  arcsec per pixel and smoothed with a 2-dimensional Gaussian of
 $\sigma = 3$ pixels.
The image includes contours from the radio map described in Section \ref{observations}. The contour levels are 0.005, 0.001, 0.05, 0.021 and 0.21 Jy beam$^{-1}$. The regions described in the text are labelled: 1 for region 1, 2 for region 2 and 3 for region 3, 4 for region 4 and 5 for region 5. The smoothing scale is chosen so as to allow both thermal and lobe related (inverse-Compton) emission to be seen. } 
\label{image}
\end{figure*}

\begin{figure*}
\epsfxsize 12.5cm
\epsfbox{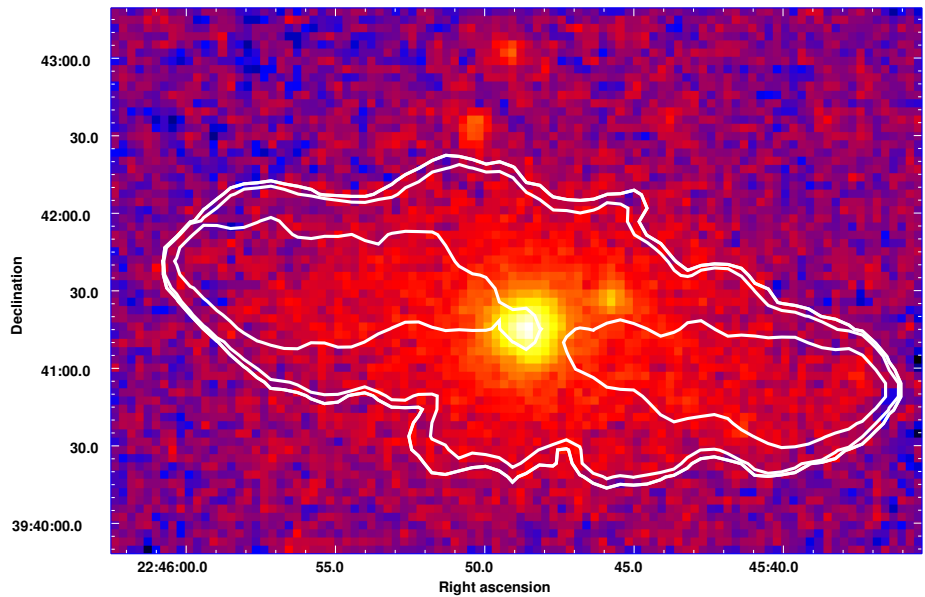}
\caption{An unsmoothed X-ray image of the AGN and the lobes of
  3C\,452 in the energy band 0.3-8.0 keV, binned to 3.2 arcsec per pixel. Radio
  contours are overlaid as in Fig.\ \ref{image}.}
\label{lobes}
\end{figure*}

\subsection{Regions for spectral extraction}
\label{spectralregion}
We extracted spectra for the following regions:
\begin{enumerate}

\item An inner region (radius 160 arcsec) surrounding the source
  (region 1 in Fig.\ \ref{image})
\item An outer annulus (radii between 160 and 320 arcsec: region 2 in
  Fig.\ \ref{image})
\item A region just outside the end of the E lobe (region 3 in
  Fig.\ \ref{image})
\item A blank-sky region for use in the double subtraction method (an
  annulus between 320 and 480 arcsec as described in section \ref{background},(region 4 in Fig.\ \ref{image}).
\item The background region used for region 3 (region 5).
\item The AGN and lobes 
\end{enumerate}

We extracted the AGN spectrum by taking a circular region with a
radius of 25 arcsec (38.25 kpc) around the AGN and used a local
annulus with a radius of 59 arcsec (90.27 kpc) for the background. We
also excluded a 12 arcsec (18.36 kpc) region to the North-East of the
AGN to remove a point source that would have affected our results. 
The same method was used for the lobes, but instead of a circular
region, we took a rectangular region of length 247 arcsec (378 kpc) by
130 arcsec (199 kpc). The regions for the AGN and the region to the
North-East of the AGN were excluded as well. We used the double
subtraction method to determine the background for this region.

Double subtraction was also used for our two large-scale regions,
intended to measure the temperature of group-scale gas which can be
detected by eye in heavily smoothed images out to 320 arcsec. Our choice of a 160-arcsec radius (taking the AGN as the centre) boundary between inner and outer
regions was intended to separate material that could have been been
directly influenced by the radio galaxy from material that should
still be in its original undisturbed state. The distance of 160 arcsec just beyond the projected length of one lobe. We masked out point
sources seen in the {\it Chandra} data for these two regions. 

Finally, region 3 was a rectangular region just outside the eastern
edge of the eastern radio lobe.
This rectangle was 111 arcsec in length and 70 arcsec in width. We
then took a local circular background region with a 30 arcsec radius.
There were no point sources in either region. This region was chosen
to allow us to search for any cluster gas that could have been shocked
by the radio source. We do not define a corresponding region on the
western edge of the source as there is a chip gap on the $pn$ detector
that covers the edge of the lobes on that side. For the western lobe we attempted to use a very similar region. However, we were only able to use the MOS cameras for the western lobe and because of this the data quality was poor and we were not able to constrain the temperature.

\subsection{Results}

In this section we describe the results of spectral fitting to each
of the regions described above. 

\subsubsection{The AGN}

The model we used was a model that has been successfully fitted for other NLRGs, consisting of a Galactic absorbed power law together with an intrinsically absorbed
power law and an iron K$\alpha$ line, which
gives a very good fit to the data. Parameters of the best-fitting model are given
in Table \ref{agnspec}. We use a fixed photon index of 2.00 for the soft power-law component following \citep{hec06}, since the photon index is poorly determined.

\begin{table*}
\caption{Best-fitting spectral parameters for the background model; abundance and redshift were fixed values.}
\label{background-fit}
\begin{tabular}{| l | l | l |}
\hline
Model & Parameter  & Value  \\
\hline
apec & kT(keV) & $0.055_{-0.003}^{+0.012}$\\
 & Normalization ($10^{14}$ cm$^{-5}$) & $3.05\pm1.95\times10^{-4}$\\
 & kT(keV) & $0.31\pm0.01$\\
 &Normalization ($10^{14}$ cm$^{-5}$) & $2.26_{-0.37}^{+0.20}\times10^{-5}$\\
zwabs &
nH ($10^{22}$cm$^{-2}$) & 0.1130\\
powerlaw & Photon Index & 1.41\\
 & Normalization (photons  keV$^{-1}$ cm$^{-2}$s$^{-1}$)  & $8.08_{-0.19}^{+0.37}\times10^{-5}$\\
\hline
{$\chi^2$/dof} & 789/468\\
\hline
\end{tabular}
\end{table*}

\begin{table*}
\caption{Best-fitting spectral parameters for the AGN}
\label{agnspec}
\hskip -2.0cm
\begin{tabular}{| l | l | l |}
\hline
Model & Parameter & Value\\
\hline
wabs &
nH ($10^{22}$cm$^{-2}$) & 0.1130 (fixed)\\
powerlaw &
Photon Index & 2.00 (fixed)\\
 & Normalization  (photons  keV$^{-1}$ cm$^{-2}$s$^{-1}$) & $1.03_{-0.26}^{+0.03}\times10^{-4}$\\
zwabs &
nH ($10^{22}$cm$^{-2}$) & $43.59_{-4.37}^{+5.27}$\\
 & Redshift & 0.0811\\[2pt]
powerlaw &
Photon index & $1.26_{-0.27}^{+0.32}$\\
 & Normalization (photons  keV$^{-1}$ cm$^{-2}$s$^{-1}$) & $6.42_{-2.19}^{+9.33}\times10^{-4}$\\
Gaussian &
E(keV) & $6.42\pm0.02$\\
 & $\sigma$ & $0.08\pm0.03$\\
 & Normalization  (photons cm$^{-2}$ s$^{-1}$) & $1.76_{-0.32}^{+0.35}\times10^{-5}$\\
\hline
Unabsorbed 0.5 - 10 keV luminosity (ergs s$^{-1}$): & $(1.02\pm0.2)\times10^{44}$\\
$\chi^2/$dof & 1252/1058\\
\hline
\end{tabular}
\end{table*}

\begin{table*}
\caption{Results from spectral analysis.}
\label{spectroanalysis}
\hskip -3.0cm
\begin{tabular}{|cccccc|}
\hline 
Region  & \multicolumn {2}{c}{power law} &\multicolumn {2}{c}{Apec} & $\chi^2$/dof \\
\hline
 & Photon Index & Normalization (photons  keV$^{-1}$ cm$^{-2}$ s$^{-1}$)  &
T(keV) & Normalization ($10^{14}$ cm$^{-5}$)\\
\hline
Lobes & $1.71_{-0.17}^{+0.13}$ & $5.76_{-1.54}^{+0.83} \times10^{-5}$ &$1.36_{-0.17}^{+0.39}$ &$5.98_{-2.77}^{+5.82}\times10^{-5}$ & 1229/1049\\
Region 1 & - & - & $1.18\pm0.11$ & $1.01_{-0.25}^{+0.24}\times10^{-4}$ & 1146/1047 \\
Region 2 & - & - & $0.86_{-0.05}^{+0.13}$ & $7.6\pm0.5\times10^{-5}$ & 862/623\\
Region 3 & - & - & $1.72_{-0.49}^{+0.9}$ & $4.56\pm1.23\times10^{-5}$ & 35/25\\
\hline
\end{tabular}
\end{table*}

\begin{figure*}
\epsfxsize=8cm
\begin{turn}{270}
\epsfbox{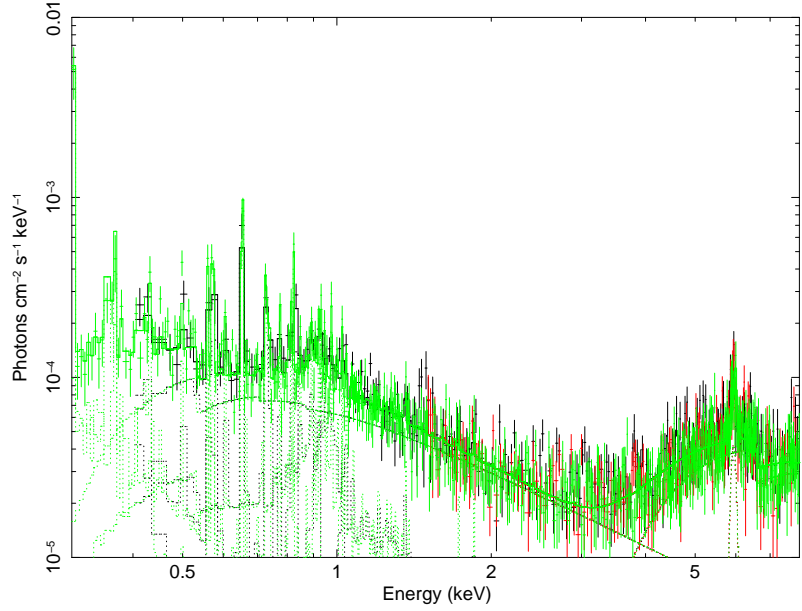}
\end{turn}
\caption{Unfolded spectrum for the AGN, lobes and environment  after particle
  background subtraction. The value for the reduced $\chi^2$ for the
  fit to this spectrum was 1.17. The iron K$\alpha$ line at 6.4 keV is
clearly visible in this plot. The green represents the PN data, the black is the MOS1 data and the red is the MOS2 data.}
\label{LS}
\end{figure*}

\begin{figure*}
\epsfxsize=8cm
\begin{turn}{270}
\epsfbox{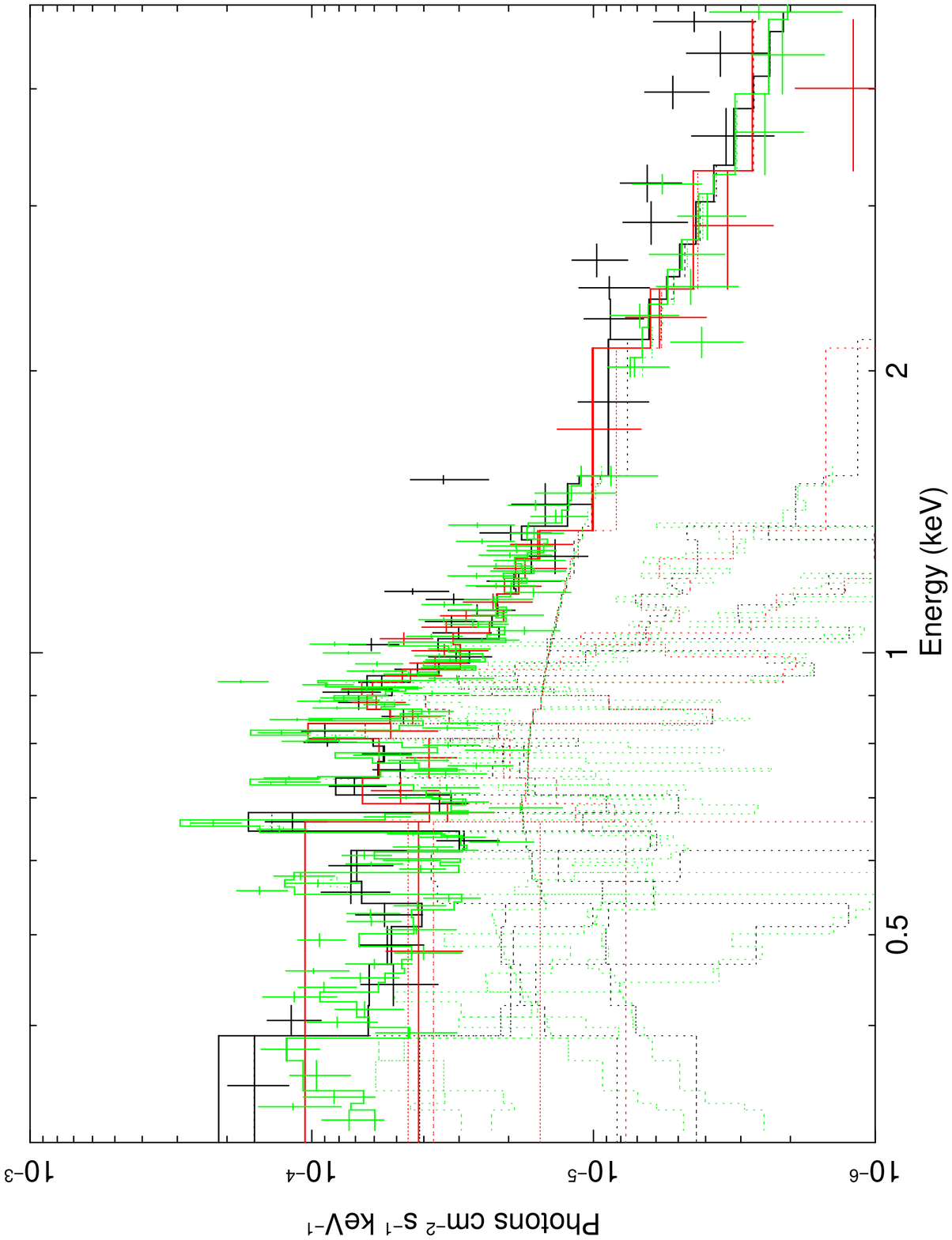}
\end{turn}
\caption{Unfolded spectrum for the outer region (region 4). This has been rebinned in {\sc xspec} to clearly show the model components. We used an absorbed apec model for fitting the data as described in Section \ref{outerregions}. The reduced $\chi^2$ for the fit to this spectrum was 1.38. The green represents the PN data, the black is the MOS1 data and the red is the MOS2 data. }
\label{ORspectra}
\end{figure*}

\subsubsection{The lobes}

To avoid having to mask out a large region, we extracted a spectrum
for both the AGN and the lobes together and then fixed the parameters
of the AGN model to the best-fitting values found in the analysis of
the AGN spectrum (see previous section). We then added to the model a
power law to model inverse-Compton emission from the lobes. In
addition, we tried the fit both with and without an {\it apec} model
intended to represent thermal emission from the regions in front of
and behind the lobes in our extraction region. We found that the fit
was marginally improved with the inclusion of an {\it apec} model with Galactic absorption and abundance
fixed at 0.3 solar
($\chi^2/$dof was 1229/1049 with  and 1241/1051  without the {\it apec} model) and so,
given that we expect some thermal emission to be present in the
extraction region and that this makes our analysis consistent with
that of \citet{itmismma02}, we chose to include it in our final fits. 
 We find a photon index for the power-law component of $1.71_{-0.17}^{+0.13}$, and its
normalization corresponds to a 1 keV flux density of
$39.9_{-4.1}^{+4.6}$ nJy. Parameters for the best-fitting models
are given in Table \ref{spectroanalysis}. For completeness, we also
attempted to model the lobes with an {\it apec} model alone, but this
gave a significantly poorer fit ($\chi^2$/dof of 1650/1051) so we do
not consider this model further. The temperature we find for the lobes is consistent within the errors with the temperature we determine for region 1.

\subsubsection{Regions outside the lobes}
\label{outerregions}

For region 1, we adopted the same approach as described above for the
lobes: the spectrum we extracted was for the whole region inside 160
arcsec and we then fixed the parameters of our best-fitting model for
the AGN and lobes in the fit. The spectrum for the AGN, lobes and the environment is shown in Figure \ref{LS}. The spectrum for the outer region (region 4) is shown in Figure \ref{ORspectra}. The model for the extended emission in
this region was an {\it apec} model with Galactic absorption and
abundance fixed to 0.3 solar; we found that a temperature of $1.18\pm 0.11$
keV provided a very good fit to the data.

For region 2 we simply fitted an {\it apec} model with Galactic absorption and
abundance fixed to 0.3 solar to the background-subtracted data. We
found a best-fitting temperature of $0.86_{-0.05}^{+0.13}$ keV.

Fitting the same {\it apec} model to the region at the edge of the
lobes (region 3) we find a best-fitting temperature of
$1.7^{+0.9}_{-0.5}$ keV. Parameters for the best-fitting models of
all three regions are given in Table \ref{spectroanalysis}. 

\section{Spatial Analysis}

\subsection {Radial Profiling}

To characterize the extended, thermal emission around 3C\,452 we
generated a radial profile of the X-ray emission in the 0.3 -- 8.0 keV
band. We binned the data adaptively to give a similar signal-to-noise
ratio in each radial bin. We masked out the lobe regions to avoid any
contribution from inverse-Compton emission to the radial profile, but the AGN was not masked out. We were able to do this by using two boxes to exclude the lobes as shown by the green boxes in Figure \ref{image}.

The particle background was modelled using the closed-filter images
described in Section \ref{background}. The files were
vignetting-corrected and filtered in the same way as the
3C\,452 observations and were rotated to be at the same angle as the
dataset using the SAS task {\it attcalc}. Using the scaling factors
described in Section \ref{background}, we were then able to subtract a
position-dependent background correction from each radial bin. The
X-ray background contribution was subtracted using the outermost
annulus of our profile (region 4 in figure \ref{image}) Finally, the radial profiles for each camera
were combined to produce a profile of net count rate per unit area.

We then fitted a model consisting of a point source, with free
normalization, and a $\beta$ model with free $\beta$, core radius and
normalization to the radial profile. Each component of the model was
convolved with an {\it XMM-Newton} point spread function derived from
the combined point-spread functions of the three cameras, suitably
weighted. The fitting process used a Markov-Chain Monte-Carlo
algorithm described in detail in \citet{gch11}, an updated
version of the method used by \citet{chbwl08}.

\begin{figure*}
\epsfxsize 8cm
\epsfbox{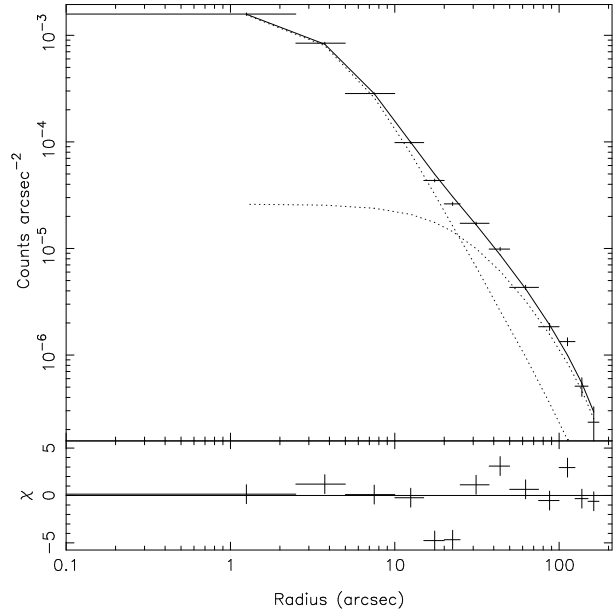}
\caption{Surface brightness profile for 3C452 with the best fitting point source + $\beta$ model which has $\beta = 0.53$ and core radius 21.98 arcsec.}
\label{profile}
\end{figure*}

The best-fitting $\beta$ model, ($\chi^2/{\rm dof} = 66/9$) has
$\beta = 0.53\pm0.02$ and the core radius was $21.98\pm1.05$ arcsec.
Figure \ref{profile} shows the surface brightness profile with the
best-fitting model. Although it would have been possible to fit more
complex models (e.g. a double $\beta$ model as used by \citealt{chhbbw05}) to the data, our aim was just to characterise the density and
pressure profiles in the inner regions and this simple model provides
an adequate representation of the data here.

The surface brightness profile presented in Figure \ref{profile} takes
into account only the statistical uncertainty in surface brightness
based on the count levels in the source and background annulus region.
It is important also to consider any systematic uncertainty introduced
by the use of the filter-wheel closed (FWC) background files in the
double-subtraction process. The particle background scaling factors
have a small associated uncertainty from the target observation and
background file high-energy count rates used to determine the scaling;
however, the dominant systematic uncertainty comes from spectral
variation of the particle background between the 10-12 keV (MOS)/12-14
keV (pn) energy range used to determine the scaling, and the 0.3-8 keV
energy range used in the profiles. \citet{rp03} investigated the
spectral properties of the particle background and report the variance
in the background level in several energy bands for a large number of
{\it XMM-Newton} observations. Based on their calculations we estimate
that the systematic uncertainty on the particle background level due
to spectral variation is $\sim 24\%$ for the pn and $\sim 10\%$ for
the MOS. The double-subtraction method used compensates almost exactly
for any discrepancy between the ``true'' particle background level in
a particular profile bin and the estimated level based on the FWC
files due to this systematic uncertainty, because the ``local''
background level we determine will include any deviation in the
particle background not accounted for by the scaling FWC component;
however, because the vignetting correction in the local background
differs slightly from that in the profile bins (by up to $20\%$) a
small systematic effect can be introduced. This effect is negligible
in most profile bins (including the outermost few bins with the lowest
surface brightness where the relative vignetting term is small), and
always $<5\%$ (pn) or $<1\%$ (MOS). We therefore conclude that the
reported statistical uncertainties on the surface brightness profile
accurately reflect the true uncertainty.

\subsection {Pressure}
\label{pressure-results}
The dynamics of 3C\,452 are determined by the relationship between the
internal (lobe) and external (thermal) pressure as a function of
position in the group.

To calculate the internal pressure we used {\sc synch} \citep{hapr98}. {\sc synch} models inverse-Compton scattering of the CMB
and synchrotron photons. The inputs to {\sc synch} are an electron
energy spectrum, the geometry of the source region (we modelled the
lobes as a cylinder with length (incorporating both lobes) 280 arcsec and radius 35 arcsec) and
synchrotron flux density measurements (we used the 178-MHz data from
the 3CRR catalogue and 1.4-GHz measurements from the image discussed
in Section \ref{observations}). The electron energy spectrum was a
broken power law, with $\delta=2$ at low energies steepening to 3 at
high energies, $\gamma_{min}$ to be 100,$\gamma_{max}$ to be $10^8$
and $\gamma_{break}$ to be 6,000. {\sc synch} fits the electron
spectrum to the radio observations and outputs an inverse-Compton
prediction for a given magnetic field strength. The magnetic field
strength is found by comparing the observed X-ray emission at 1 keV
and the predicted X-ray inverse-Compton emission; we measure a lobe
magnetic field strength of $0.175 \pm 0.010$ nT, significantly below
the equipartition value of 0.52 nT. Assuming that only the
relativistic electrons and the magnetic field contribute to the lobe
pressure (\citealt{cbhw04} have tested this assumption) then the
electron spectrum and magnetic field strength give us a total energy
density and internal pressure in the lobes. We found the internal
pressure to be $2.6\times10^{-13}$ Pa (the systematic uncertainties dominate over the statistical uncertainties and for this reason we present the internal lobe pressure without errors).

\begin{figure*}
\epsfxsize 8cm
\epsfbox{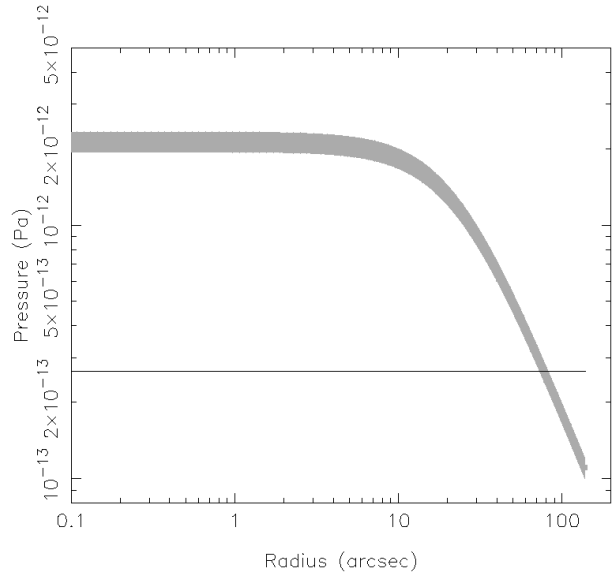}
\caption{Pressure profile which shows the external pressure as a
  function of radius. The straight line across represents the internal
  pressure. Where the external pressure lies above the line, the lobes are contracting; below the line, the lobes are expanding. }
\label{Pressure}
\end{figure*}

From the surface brightness profile and the measured temperature of
the external gas we can find the external pressure at a given radius
from the AGN, using the method of \citet{bw93}.
Fig.\ \ref{Pressure} shows the pressure as a function of radius.
We see that the internal pressure is less than the external pressure
at radii less than $\sim 80$ arcsec; thus the source should be
contracting in its inner regions. Further out, though, the source is
overpressured, and the difference is greatest at the far end of the
lobes, where $p_{\rm int}/p_{\rm ext} = 2.4$, high enough that the
source should be driving a shock into the external medium. 

Projection affects both the size of the source (thus affecting the
inferred internal pressure from the inverse-Compton measurements) and
the environment in which it is embedded, so in general the effect of
projection on measurements of internal to external pressure ratio is
complicated (e.g. \citealt{hw00}). In this particular case
we expect that the source will be no more than $45^\circ$ from the
plane of the sky, since it is a narrow-line radio galaxy and the
unification angle for these objects is of that order (e.g. \citealt{b89}; \citealt{hwb98}). Investigating different projection
angles in the range $0^\circ < \theta < 45^\circ$, we find that
although an increase in $\theta$ means a drop in internal pressure, as
expected, it tends to increase the ratio between the internal and
external pressures, up to a value of 3.2 at $\theta = 45^\circ$: so if
anything the expected shock would be stronger, and our conclusions
about the source dynamics are unchanged by plausible projection factors.

We discuss the implications of these results in Section
\ref{discussion}.

\section{Discussion}
\label{discussion}

\subsection{AGN}

The model that best fits our {\it XMM-Newton} data gives us an
unabsorbed 0.5-10 keV luminosity for all components of the model of
$(1.44^{+0.08}_{-0.07})\times10^{44}$ ergs s$^{-1}$. This is not
directly comparable to the results of the {\it Chandra} analysis
presented by \citet{itmismma02} and \citet{ewhkb06}, since they
used a slightly different model. To make a direct comparison and
search for any evidence of AGN variability, we used the same
extraction regions that we used for the AGN spectrum on the {\it Chandra}
data used by \citet{itmismma02} and \citet{ewhkb06}. We found the
unabsorbed 0.5-10 keV luminosity for the {\it Chandra} AGN spectrum to be
$(1.55^{+0.24}_{-0.19})\times10^{44}$ ergs s$^{-1}$. We directly
compared key parameters of both datasets (e.g.\ the absorbed column
density) and found that there were no significant differences between
the {\it Chandra} and {\it XMM-Newton} data. This suggests that there
has been very little variability in the 6 years between the
observations using {\it Chandra} and {\it XMM-Newton}.

\subsection{Lobes}

Inverse-Compton emission from the lobes is clearly detected in our
{\it XMM-Newton} data. Our flux density measurement ($40 \pm 4$ nJy at
1 keV) is in excellent agreement with the results from {\it Chandra}
data: \citet{itmismma02} give $41 \pm 15$ nJy while \citet{chbw05} , in their analysis of the same data, quote $37 \pm 2$
nJy. Both previous analyses, like ours, fitted a combination of thermal
and power-law models to the spectrum of the lobe region.  The
differences in the error bars presumably arises in part because Isobe
\etal\ allowed the Galactic $N_{\rm H}$ to vary; we follow \citet{chhbbw05}
in keeping it fixed. Our measurement of the photon index of the
power-law component is also very consistent with the results of
\citet{chbw05}.

\subsection{Environmental impact}

We can make a rough estimate of the work done by the lobes on the
environment. Modelling the lobes as a cylinder in the plane of the sky
(as described above, Section \ref{pressure-results}) the total lobe
volume is $1.15\times10^{65}$ m$^3$. The energy stored in the lobes is
then the energy density times the volume, or $9.1 \times 10^{52}$ J.
Roughly equating the work done on the environment with 1/3 of this, we
obtain $PV \approx 3.0\times10^{52}$ J or $3.0\times10^{59}$ ergs.

As we saw in Section \ref{outerregions}, the outer regions of the host
group are cooler by $0.32 \pm 0.14$ keV than the regions at radii
comparable to the extent of the lobes. We can use our estimate of the
$PV$ work done by the lobes to ask whether this temperature difference
is consistent with the idea that the radio source is heating the IGM
in the inner part of the group. 

Using the standard formulae for the proton number density in a $\beta$
model and the relation between the number of electrons and the
number of protons (e.g. \citealt{bw93}), we can calculate
the thermal energy stored in a sphere of gas within the region that
could plausibly have been affected by the radio source (region 1).

\begin{equation}
n_p=n_{p0}\left(1+\frac{r^2}{r_{cx}^2}\right)^{\frac{-3\ \beta}{2}}
\label{betamodel1}
\end{equation}
\begin{equation}
n_e=1.18n_p
\end{equation}
where
$n_{p0}$ is the central proton number density which we obtain from the pressure profile ( Fig.\ref{Pressure}) ,
$\beta$ and $r_{cx}$ are the parameters of the $\beta$-model found by
fitting to the surface brightness profile, and $n_p$ and $n_e$ are the proton and electron number densities at a given radius.

We then integrate equation \ref{betamodel1} out to $r=245$ kpc (160
arcsec) to determine the total number of particles in our sphere of
influence, $N_{\rm tot} = 3.24^{+0.44}_{-1.07}\times10^{68}$; multiplying this by ${3\over 2}kT$ (where $T$ is the temperature of the gas in region 1 as found in the spectral extraction) gives the thermal energy
stored in the region, $9.17_{-2.93}^{+0.89}\times10^{52}$ J or
$9.17_{-2.93}^{+0.89}\times10^{59}$ erg. Effectively, this calculation
also gives us the heat capacity of the gas in the region, on the
assumption that all work done on the gas remains present as excess
heat, $C = {3\over 2}kN_{\rm tot}$. Using this, we find that the work
done by the AGN (as calculated above) should increase the temperature
of the inner region by 0.39 keV. This is very consistent with the
observed higher temperature in the centre of the group. There are
several important caveats in doing this simple comparison: firstly,
the estimate of $PV$ work done is probably not accurate to within
better than a factor of 2; secondly, we might expect that at least
some of the work done by the radio source will go into lifting gas out
of the centre of the group, i.e. into potential and not thermal
energy; and thirdly, we do not know the original (undisturbed)
temperature of the gas in region 1, and observations of groups and
clusters show that the temperature profiles are unlikely to be flat
(e.g. \citealt{pbcabft07}). However, it would obviously be possible for
the estimate of AGN work done and the temperature difference to be
grossly inconsistent (e.g. if the central temperature had been
significantly less than the outer temperature) and this is not what we
observe.

Finally, we note that the energy stored in the lobes is very
comparable to the {\it total} thermal energy in region 1, and thus
comparable to the gravitational binding energy of a significant
fraction of the mass in the group. If the energy stored in the lobes
is eventually thermalized, it will have a much more dramatic effect on the
intra-group gas than the current work being done by the radio source.

\subsection{Shock heating}

Our results show that the ratio between the internal pressure and the
external pressure is greater than 2 at the edge of the lobes. This
suggests that the lobes should be expanding supersonically. We can
estimate the temperature of the post-shocked gas in order to compare
it to what is seen observationally. Using the Rankine-Hugoniot
relationships, we can derive equations for the pressure and
temperature in terms of the Mach number and the polytropic index. For the pressure,
\begin{equation}
\frac{P_2}{P_1}= \frac{2\Gamma M_1^2+(1-\Gamma)}{\Gamma+1}
\label{Pressureeq}
\end{equation}
where $P_2$ is $2.64\times10^{-13}$ Pa (i.e. we assume that the
post-shock pressure is equal to the lobe pressure), $P_1$ is
$1.11\times10^{-13}$ Pa (the external pressure)
and $\Gamma$ is $\frac{5}{3}$. Rearranging for $M_1$, we find that the Mach number is 1.6.

We can then determine the expected temperature using the relation
between pre- and post-shock temperatures:
\begin{equation}
\frac{T_2}{T_1}= \frac{[2\Gamma M_1^2+(1-\Gamma)][\Gamma-1+2M_1^2]}{(\Gamma+1)^2}
\label{tempeq}
\end{equation}
where $T_1$ is 1.18 keV (pre-shocked gas temperature), $\Gamma$ is
$\frac{5}{3}$ and $M_1$ is 1.6. We find the post-shocked gas temperature
($T_2$) to be 1.94 keV. In Section \ref{outerregions}, we found that
the temperature in the region we defined at the edge of the lobes
(region 3) was $1.72_{-0.49}^{+0.9}$ keV, which is at least consistent
with the idea that the lobes are driving a shock of the expected
strength into the IGM at their overpressured tips.  However, there is
no direct evidence for such a shock in the surface brightness profiles
-- unsurprisingly, since the density change in a weak shock like this
is small. Much deeper observations will be needed to detect shocks
around sources like 3C\,452 if, as our modelling suggests, they are
confined to the outer few tens of kpc of the lobes.

\section{Summary and conclusion}

We have presented a deep {\it XMM-Newton} observation of the AGN,
lobes and group environment of the nearby, powerful FRII radio galaxy
3C\,452. This is the most sensitive X-ray observation of this source
made so far and it has allowed us to measure the X-ray properties of
the source more accurately than has previously been possible. We
have compared the results from {\it XMM-Newton} with those derived
from an earlier {\it Chandra} observation, and shown that there is
very little variation in the AGN properties; there is no sign of
variability in this object. Our data give an excellent detection of
the inverse-Compton emission from the lobes; the flux density  we
measure is very consistent with earlier work.
 We have used our measurements of the properties of the group
environment, possible for the first time with the {\it XMM-Newton}
observations, to search for any large scale heating caused by the
radio source in the environment of 3C\,452. We measured the temperatures of three regions. Region 1 is a region with a radius of 160 arcsec, centred on the AGN with a temperature of $1.18\pm0.11$ keV. Region 2 has a radius of 320 arcsec, centered on the AGN with a temperature of $0.86_{-0.05}^{+0.13}$ keV. Finally, region 3 is a box of length 111 arcsec and width 70 arcsec intended to assess the temperature of any shock at the leading edge of the eastern lobe and a temperature of $1.72_{-0.49}^{+0.89}$ keV was found. Using the inverse-Compton
detection of the lobes ($40\pm4$ nJy), we have measured the internal lobe pressure,
and from that have been able to estimate the work done by 3C\,452 on
its environment. If all of this work were put into heating the
group-scale gas within a sphere defined by the size of the radio
source, we would see an increase in temperature in the central regions
of the group which is in fact very consistent with what we observe.
Our pressure measurements ($2.64\times10^{-13}$ Pa for the internal pressure and $1.11\times10^{-13}$ Pa for the external pressure (at the edge of the lobes)) also imply that the outer edges of the lobes
are overpressured and driving a shock, while the inner parts of the
lobes are underpressured and contracting; temperature measurements at
the edge of the lobes are consistent with a shock model but our data
are still not deep enough to claim a detection. Finally, we have shown
that the total energy in the lobes is comparable to the thermal energy
in the surrounding environment (the thermal energy stored in the lobes is $9.17_{-2.93}^{+0.89}\times10^{52}$ J or $9.17_{-2.93}^{+0.89}\times10^{59}$ erg), so that if the energy in the lobes is
ever thermalized it will have a much bigger effect on the environment
than the radio source is having at the present time.

\section*{Acknowledgements}
DLS would like to thank friends and family for their unwavering support throughout this work.
MJH would like to thank the Royal Society for a research fellowship.We would like to thank Etienne Pointecouteau, Monique Arnaud and Gabriel Pratt for developing the background calibration method used in this paper and for providing filter-wheel closed background datasets.
This work is based on observations obtained with {\it XMM-Newton}, an ESA science mission with instruments and contributions directly funded by ESA Member States and the USA (NASA).

\newpage \bibliographystyle{aa}
\bibliography{3c452}
\end{document}